\documentclass[prb,twocolumn,longbibliography,
superscriptaddress,
 amsmath,amssymb,
 aps,floatfix
]{revtex4-1}
\usepackage{gensymb}
\usepackage{graphicx}
\usepackage{dcolumn}
\usepackage{bm}
\usepackage{multirow}
\usepackage{textcomp, gensymb}
\usepackage{
}

\begin{document}

\title{Spectral evidence for NiPS\textsubscript{3} as a Mott-Hubbard insulator}

\author{Yifeng Cao}
\affiliation{Department of Physics, Boston University, Boston, Massachusetts, 02215, USA.}
\affiliation{Advanced Light Source, Lawrence Berkeley National Laboratory, Berkeley, California, 94720, USA.}

\author{Nicholas Russo}
\affiliation{Department of Physics, Boston University, Boston, Massachusetts, 02215, USA.}
\affiliation{Advanced Light Source, Lawrence Berkeley National Laboratory, Berkeley, California, 94720, USA.}

\author{Qishuo Tan}
\affiliation{Department of Chemistry, Boston University, Boston, Massachusetts, 02215, USA.}

\author{Xi Ling}
\affiliation{Department of Chemistry, Boston University, Boston, Massachusetts, 02215, USA.}

\author{Jinghua Guo}
\affiliation{Advanced Light Source, Lawrence Berkeley National Laboratory, Berkeley, California, 94720, USA.}

\author{Yi-de Chuang}
\affiliation{Advanced Light Source, Lawrence Berkeley National Laboratory, Berkeley, California, 94720, USA.}

\author{Kevin E. Smith}
\email{ksmith@bu.edu}
\affiliation{Department of Physics, Boston University, Boston, Massachusetts, 02215, USA.}


\begin{abstract}
The layered van der Waals trichalcogenide NiPS\textsubscript{3} has attracted widespread attention due to its unique optical, magnetic, and electronic properties. The complexity of NiPS\textsubscript{3} itself, however, has also led to ongoing debates regarding its characteristics such as the existence of self-doped ligand holes. In this study, X-ray absorption spectroscopy and resonant inelastic X-ray scattering have been applied to investigate the electronic structure of NiPS\textsubscript{3}. With the aid of theoretical calculations using the charge-transfer multiplet model, we provide experimental evidence for NiPS\textsubscript{3} being a Mott-Hubbard insulator rather than a charge-transfer insulator. Moreover, we explain why some previous XAS studies have concluded that NiPS\textsubscript{3} is a charge-transfer insulator by comparing surface and bulk sensitive spectra. 

\end{abstract}

\maketitle

\section{Introduction}
\begin{figure}[htb]
\centering
\includegraphics[width=7cm]{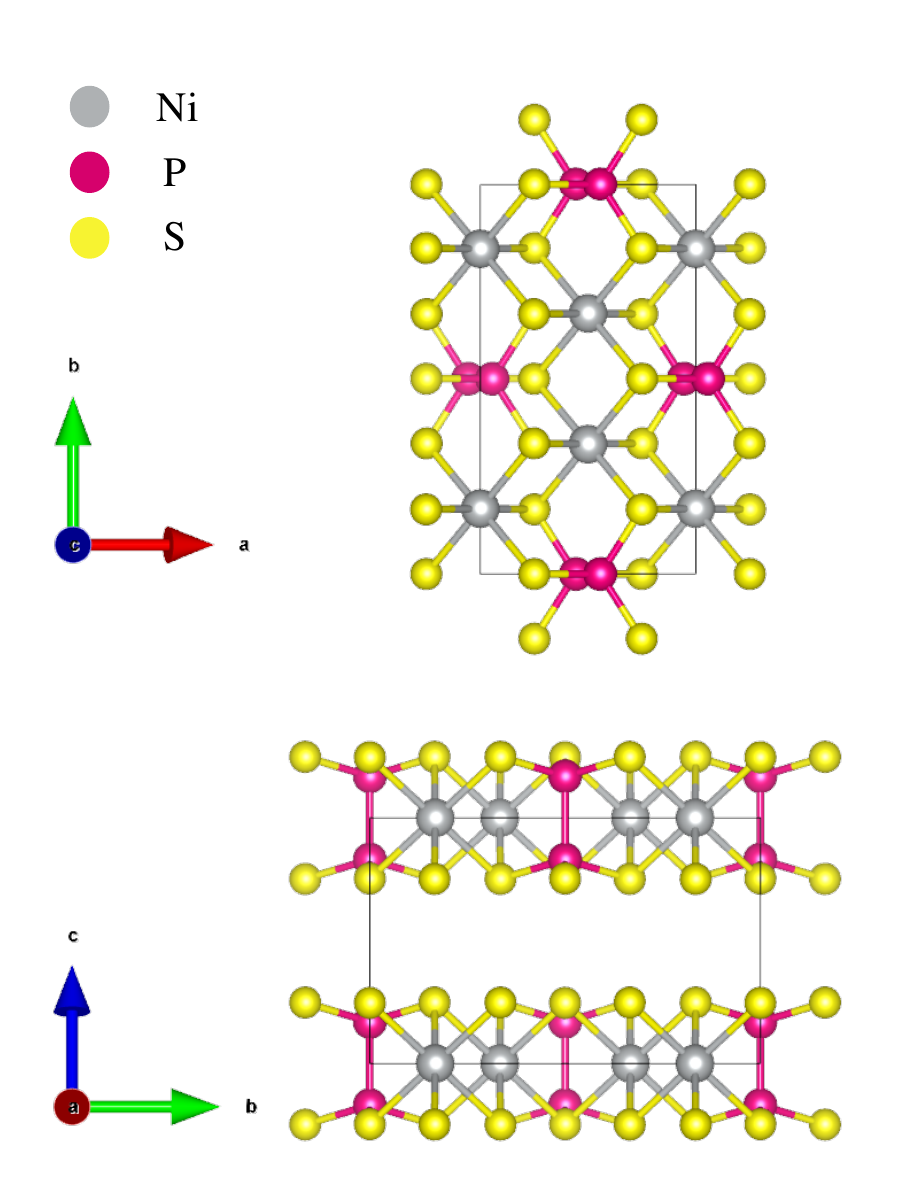}
\caption{The crystal structure of NiPS\textsubscript{3} seen along the $c$ axis (top) and $a$ axis (bottom). Generated by VESTA\cite{klingen1973uber}.}
\label{fig:1}
\end{figure}

Van der Waals (vdW) materials, including graphene\cite{novoselov2012roadmap}, transition metal dichalcogenides\cite{nicolosi2013liquid,xu2015discovery}, and boron nitride\cite{watanabe2004direct} are a promiment class of two-dimensional (2D) crystals. The reduced dimensionality of these systems often leads to exceptional physical properties, resulting from enhanced quantum effects and increased correlations due to the reduction of available phase space and diminished screening\cite{ajayan2016two}. In this study, we focus on NiPS\textsubscript{3}, a type of 2D vdW metal phosphorus trichalcogenides (MPX\textsubscript{3}). NiPS\textsubscript{3} is monoclinic, with space group $C2/m$; it can be visualized as layered NiS\textsubscript{2} crystals with P-P pairs substituting for one third of the Ni sites\cite{wang2018new,klingen1973uber} as shown in FIG.~\ref{fig:1}. NiPS\textsubscript{3} exhibits intriguing properties in electronics\cite{kim2018charge,lane2020thickness,tan2022charge},spintronics\cite{wang2021spin,basnet2021highly,scheie2023spin}, magnetism\cite{wildes2015magnetic,kim2019suppression,He2024}, and optics\cite{liu2019nips,wang2022electronic}. NiPS\textsubscript{3} stands out among the 2D crystals due to its ultrasharp photoluminescence (PL) peak at 1.47 eV, induced by a spin-orbit-entangled exciton state\cite{kang2020coherent,wang2021spin,hwangbo2021highly}. Understanding the properties of the exciton state requires an investigation of both the occupied and unoccupied states of NiPS\textsubscript{3}; although the exciton of NiPS\textsubscript{3} has been extensively studied\cite{belvin2021exciton,ho2021band,klaproth2023origin,kim2023anisotropic}, many fundamental electronic properties of NiPS\textsubscript{3} have still remained controversial.

\begin{figure}[htb]
\centering
\includegraphics[width=7.5cm]{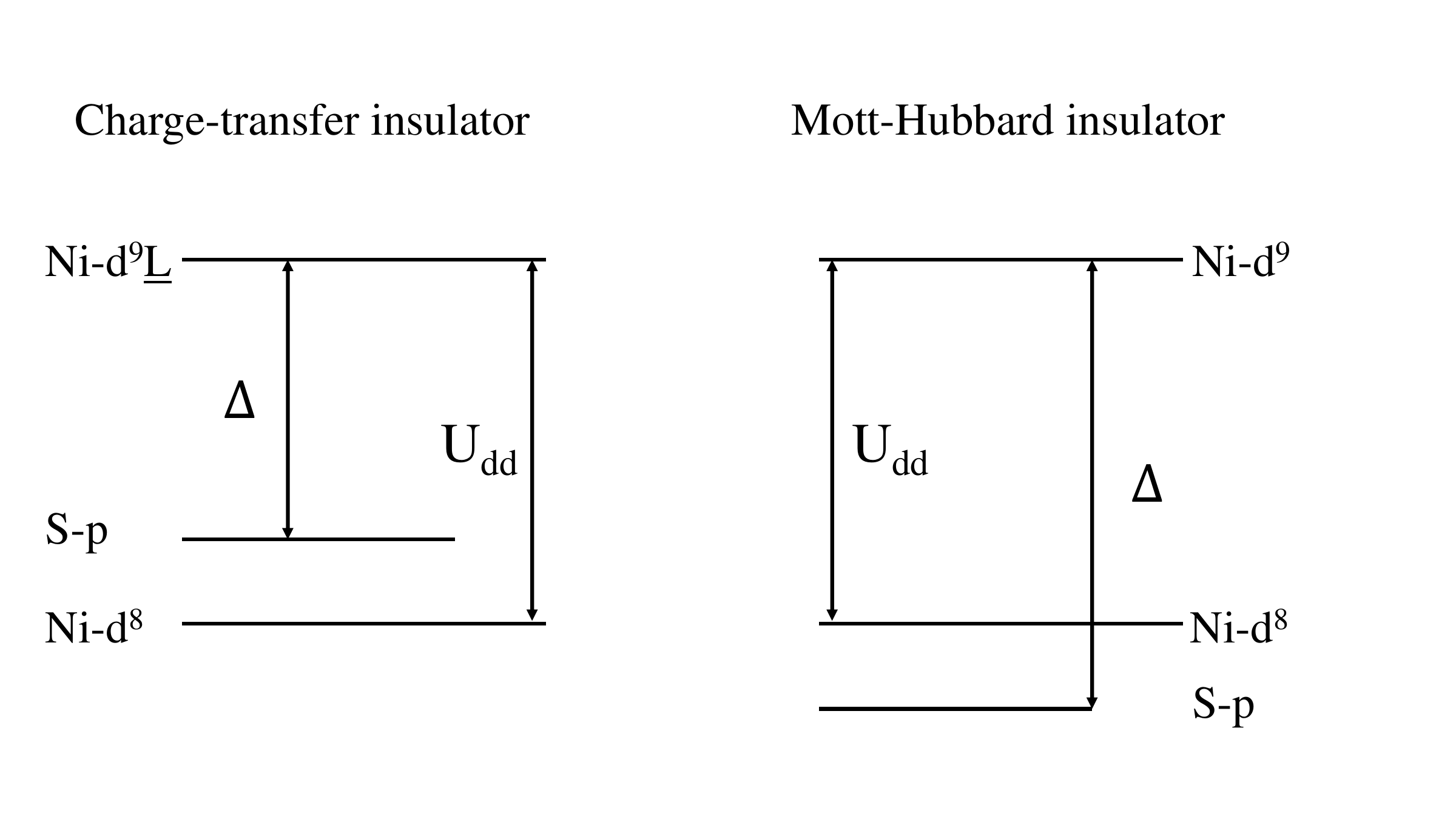}
\caption{Simplified electronic energy levels for NiPS\textsubscript{3} assuming it is a charge-transfer insulator (left) and assuming it is a Mott-Hubbard insulator (right)\cite{sahiner1996electronic}.}
\label{fig:2}
\end{figure}

In X-ray absorption spectroscopy (XAS), a core electron is excited to an empty state, thereby probing the unoccupied states of the system, with information about the occupied states being inferred indirectly\cite{fuggle1992unoccupied}. Meanwhile, in resonant inelastic X-ray scattering (RIXS), photons are scattered inelastically off matter, and we measure the excitations corresponding to the changes that occur during the scattering\cite{ament2011resonant}. Both XAS\cite{kim2018charge,kim2019mott,kang2020coherent,yan2021correlations} and RIXS\cite{kang2020coherent,discala2023dimensionality,He2024} have been applied to NiPS\textsubscript{3}, yet they have yielded differing results. The major controversy is whether NiPS\textsubscript{3} is a charge transfer insulator or Mott-Hubbard insulator. A charge-transfer insulator is defined by a charge-transfer energy $\Delta$ (the energy cost of transferring an electron from the $p$ level of a ligand to the metal ion) that is smaller than the Coulomb repulsion energy between two $d$ electrons $U_{dd}$ (the energy cost of a $d$ electron hopping to another already occupied $d$ site); by contrast, when $\Delta$ is greater than $U_{dd}$, the standard Hubbard model will be applied, resulting in a Mott-Hubbard insulator\cite{khomskii2014transition}. The simplified energy levels for these two types of insulators\cite{sahiner1996electronic} are shown in FIG.~\ref{fig:2}: although both have a dominant ground state of $d^8$, the Mott-Hubbard insulator's lowest excited state is $d^9$, while the charge-transfer insulator's lowest excited state is $d^9\underline L$, where $\underline L$ represents a ligand hole. Kim et al. first applied XAS to the Ni L-edge of NiPS\textsubscript{3} and compared with calculated spectra corresponding to different values of $\Delta$\cite{kim2018charge}. They found that NiPS\textsubscript{3} is a self-doped charge transfer insulator with a negative value of $\Delta$. With a negative $\Delta$, the charge transfer from the ligand will occur spontaneously, resulting in a dominant ground state $d^9\underline L$. This claim, however, was directly challenged by a subsequent calculation which showed that NiPS\textsubscript{3} would be a metal rather than an insulator with $d^9\underline L$ ground state\cite{kim2019mott}. They confirmed the presence of the paramagnetic Mott phase and the dominant ground state $d^8$ for NiPS\textsubscript{3}. Following XAS studies\cite{kang2020coherent,yan2021correlations}, although challenging NiPS\textsubscript{3} as having negative $\Delta$ with self-doped behavior, they still concluded that NiPS\textsubscript{3} has a small but positive $\Delta$ and belongs to the category of charge-transfer insulator. The exciton of NiPS\textsubscript{3} probed by RIXS is also a subject of debate. Initially, the exciton has been described as a Zhang-Rice singlet\cite{kang2020coherent}. However, this does not explain the extremely narrow PL peak of the exciton\cite{He2024} and its mismatch in response to an applied magnetic field\cite{jana2023magnon}. Instead of the Zhang-Rice mode, a most recent RIXS studies have revealed the dominant Hund's character in NiPS\textsubscript{3}\cite{He2024}.

In this study, we compared XAS measurements of an as-grown single crystal of NiPS\textsubscript{3} to an exfoliated sample of NiPS\textsubscript{3} in order to reconcile the controversy regarding the value of $\Delta$ for NiPS\textsubscript{3}. XAS was applied using both total electron yield (TEY) mode and total fluourescence yield (TFY) mode for surface sensitive ($<10$ nm) and bulk sensitive ($>100$ nm) measurements, respectively\cite{yang2021situ}. Therefore, the surface and bulk features of NiPS\textsubscript{3} were simultaneously collected, enabling us to capture the most subtle details of the electronic structure. By comparing the XAS results in TEY mode and TFY mode, we obtained different spectra from previous XAS studies\cite{kim2018charge,kang2020coherent,yan2021correlations} on NiPS\textsubscript{3}. Furthermore, by comparison with calculations using the charge-transfer multiplet model we are led to conclude that, contrary to the previously held belief that NiPS\textsubscript{3} has a very small value of $\Delta$, it actually has a relatively large value of $\Delta$. Further analysis was conducted to explain why previous XAS results concluded that NiPS\textsubscript{3} belongs to the charge-transfer insulator. Moreover, we also applied RIXS to NiPS\textsubscript{3} and found that $dd$ excitations are dominant, while charge-transfer excitations were too weak to be observed. Combined with the conclusion from XAS that NiPS\textsubscript{3} has a relatively large $\Delta$, we provide strong evidence that NiPS\textsubscript{3} is a Mott-Hubbard insulator.

\section{Methodology}
NiPS\textsubscript{3} single crystals were synthesized by the chemical vapor transport method. Pure elements, in a stoichiometric ratio of Ni:P:S $=$ 1:1:3 (2 g in total), along with 40 mg iodine as transport agent, were enclosed in quartz ampules under vacuum of 1$\times$10\textsuperscript{$-$4} Torr. Subsequently, the ampules underwent heating in a two-zone furnace with the temperature range of 650 $-$ 600 $^{\circ}$C for 1 week, followed by cooling to room temperature. Bulk crystals were harvested from the lower temperature zone of the ampules. The as-grown NiPS\textsubscript{3} single crystal was directly mounted on the sample holder with conductive copper tape. The exfoliated sample was prepared by removing the top layers using Scotch tape under an argon atmosphere in a glovebox, and was transferred to the analysis chamber using a vacuum suitcase to prevent oxidation. 

The nickel L\textsubscript{3,2}-edge XAS spectra were measured at Advanced Light Source (ALS) beamline 7.3.1 at Lawrence Berkeley National Laboratory (LBNL). Both TEY and TFY modes were collected simultaneously. The XAS photon energy was calibrated using a Li(Ni\textsubscript{1/3}Mn\textsubscript{1/3}Co\textsubscript{1/3})O\textsubscript{2}  reference sample. O K-edge XAS spectra with both TEY and TFY modes were also collected to determine the extent of O contamination in the sample.

The theoretically calculated XAS spectra were obtained using the Crispy software for the Quanty library\cite{retegan_crispy}. A charge-transfer multiplet model is applied to the system, described by a semi-empirical Hamiltonian that includes atomic and crystal-field interactions\cite{de20212p}. In this study, nickel L\textsubscript{3,2}-edge XAS spectra were calculated by Crispy for various values of $\Delta$. For simplicity, the symmetry is set to $O_h$. The applied parameters are: $F_k$=0.8, $G_k$=0.8, $\zeta$=1.0, $U(3d,3d)$=4.0 eV, $U(2p,3d)$=6.0 eV, $10Dq(3d)$=1.0 eV, $Ve_g(3d, L1)$=1.4 eV, $Vt_{2g}(3d,L1)$=1.0 eV. The broadening full width at half maximum parameters (eV) are set to be 0.48, 0.52 for Lorentzian and 0.3 for Gaussian. The values of $\Delta$ range from -1 eV to 5 eV.  

The RIXS spectra of nickel L\textsubscript{3}-edge were obtained at beamline 8.0.1 at the ALS. The RIXS photon energy was calibrated by an undulator.

\section{Results and discussion}

\begin{figure}[htb]
\centering
\includegraphics[width=7cm]{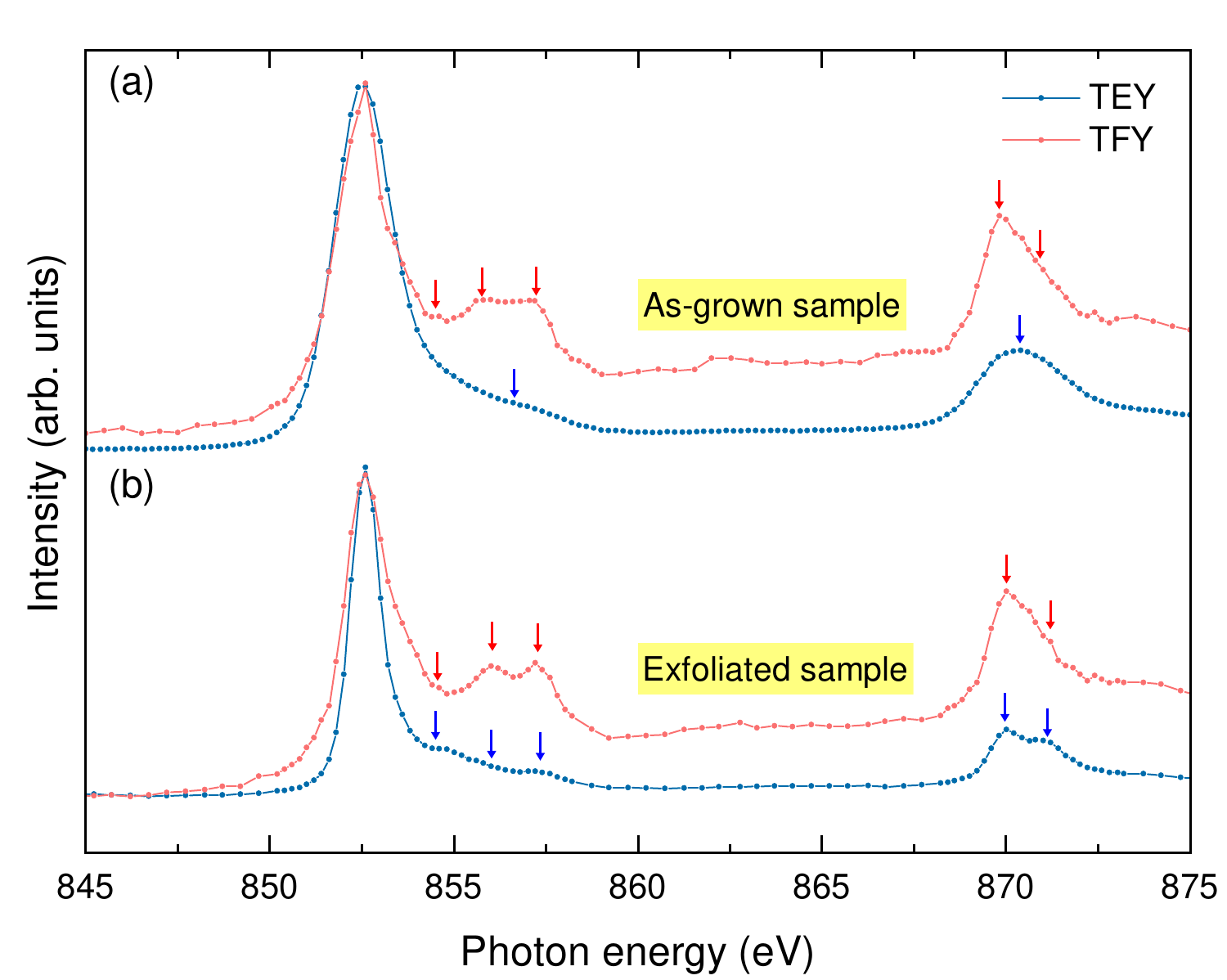}
\caption{Nickel L\textsubscript{3,2}-edge XAS spectra with TEY mode (blue) and TFY mode (red) for (a) an untreated NiPS\textsubscript{3} sample and (b) a NiPS\textsubscript{3} sample after exfoliation of the surface layers in an air-isolated environment.}
\label{fig:3}
\end{figure}

XAS was first applied to nickel L\textsubscript{3,2}-edge of as-grown NiPS\textsubscript{3} single crystal as shown in FIG.~\ref{fig:3}(a). Surprisingly, the spectrum from TEY mode (blue solid line) and TFY mode (red solid line) exhibited different features. In TEY mode, both the L\textsubscript{3}-edge and L\textsubscript{2}-edge show a single peak feature with a satellite peak indicated by the blue arrow around 856.6 eV. This is precisely the feature corresponding to the small $\Delta$ observed in previous literature through experiments and calculations, indicating a charge-transfer insulator. However, as mentioned earlier, the TEY signal primarily reflects the surface electronic structure of a sample. In contrast, the TFY spectra reveal that the satellite splits into two peaks at 855.8 eV and 857.2 eV. Furthermore, distinguishable double peak features appear at both L\textsubscript{3}-edge (854.4 eV) and L\textsubscript{2}-edge (871.0 eV), as indicated by the red arrows.

\begin{figure}[htb]
\centering
\includegraphics[width=7cm]{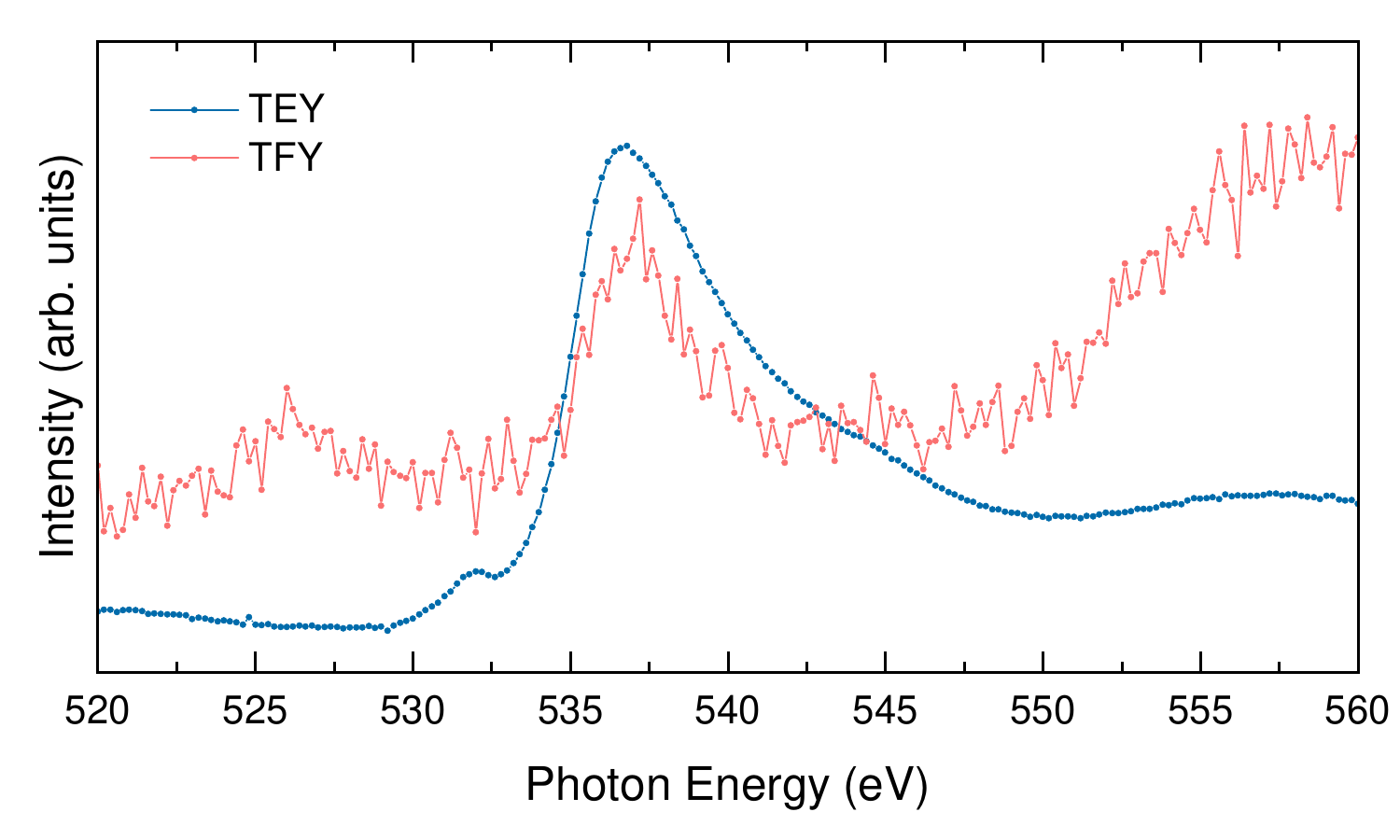}
\caption{Surface oxidation check: O K-edge XAS spectra with TEY mode (blue) and TFY mode (red) for the untreated NiPS\textsubscript{3} sample used in FIG.~\ref{fig:3}(a).}
\label{fig:4}
\end{figure}

Although many studies indicate that NiPS\textsubscript{3} is stable in air\cite{lu2020exfoliation,fang2021situ}, exposure to air may still cause surface oxidation and affect its TEY signal\cite{frati2020oxygen}. This can be inferred from the different features exhibited by the TEY and TFY signals in FIG.~\ref{fig:3}(a). To chracterize the surface oxidation, we also applied XAS in the oxygen K-edge region for the same sample used in FIG.~\ref{fig:3}(a) as presented in FIG.~\ref{fig:4}. As we previously inferred, the surface-sensitive TEY signal in the oxygen range exhibited very distinct signals, with characteristic peaks appearing in both the X-Ray absorption near edge structure (XANES) region (532 eV and 537 eV) and extended X-ray absorption fine structure (EXAFS) region (558 eV). This indicates that the oxygen on the surface of NiPS\textsubscript{3} is not merely adsorbed free oxygen from the air, but has chemically bonded with the Ni cations. The spectrum in TFY mode, on the other hand, shows a very weak signal-to-noise ratio (SNR) in oxygen range compared to the spectrum in TEY mode. The appearance of an oxygen peak in the bulk-sensitive TFY mode does not mean that the entire NiPS\textsubscript{3} sample is oxidized. Instead, it is because the TFY signal detection range covers the entire region to a depth of about 100 nanometers, including the few nanometers of the oxidized surface. It is precisely because the SNR of the TFY signal is significantly lower than that of the TEY signal that it proves the oxidation is only present on the surface of the NiPS\textsubscript{3} sample (occupying only a small portion of the TFY detection depth). This is reasonable, as the bulk NiPS\textsubscript{3} should not retain oxygen and should be air-stable.

Further investigation into the structure of the O K edge XAS has found that it is likely that nanoscaled NiO appear on the surface of NiPS\textsubscript{3}, evidenced by the similarity to the O K edge XAS from NiO nanowires as shown by Wu \textit{et al.}\cite{wu2005structural}; these nanoscaled NiO exhibit very different features in the XAS O K-edge region compared to bulk NiO. By comparing with their XAS results, we find that the characteristic peaks in the TEY mode in FIG.~\ref{fig:4} are perfectly consistent with those of NiO nanowires in both the XANES and EXAFS regions. The TEY probe depth shows that the nanoscaled NiO once again demonstrates that oxidation occurs only on the surface of the NiPS\textsubscript{3} sample.

To eliminate the impact of surface oxidation on the experimental results, the most straightforward method is to directly remove the outermost few layers. As a vdW material, NiPS\textsubscript{3} can be easily exfoliated using Scotch tape. The XAS spectrum was remeasured for the Ni L\textsubscript{3,2}-edge of exfoliated NiPS\textsubscript{3} sample [FIG.~\ref{fig:3}(b)]. After exfoliation, we observed the following: (1) The TEY and TFY spectra exhibited excellent consistency, as indicated by the blue and red arrows, respectively; compared to the untreated sample, (2) the TEY signal showed significant changes, with the most notable being that both the Ni L\textsubscript{2} peak and L\textsubscript{3} peak split into a double peak structure, with more distinct features appearing in the satellite peak next to the L\textsubscript{3} peak; (3) the TFY signal did not show obvious changes. These observations indicate that: (1) Exfoliation successfully removed the surface oxidation layers, making the TEY signal after exfoliation more likely to reflect the true sample signal; (2) The presence of the surface oxidation layers have a significant impact on the TEY signal; (3) The oxidation layers have a minimal impact on the TFY signal, allowing the TFY signal to remain reliable throughout the entire study.

\begin{figure}[htb]
\centering
\includegraphics[width=7cm]{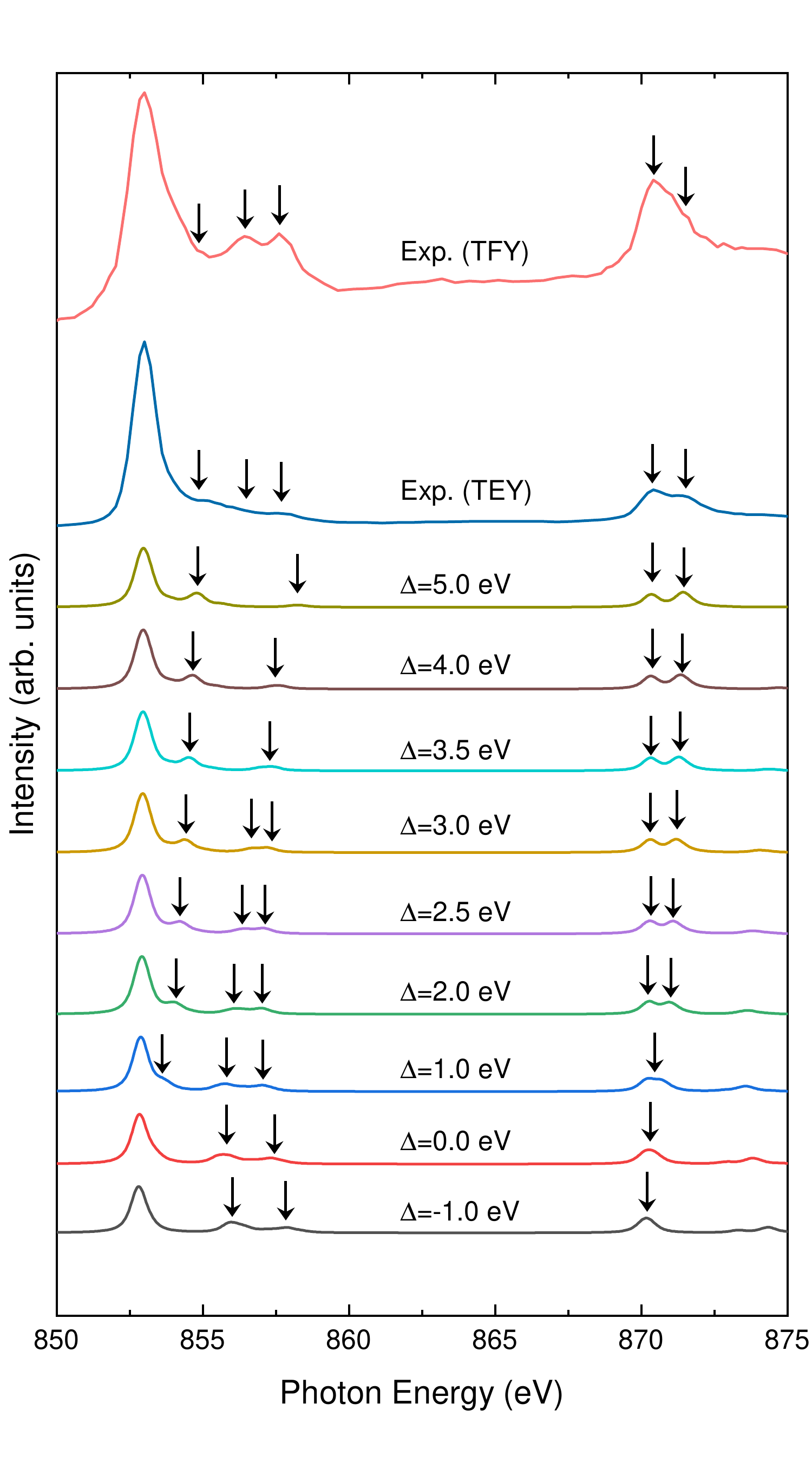}
\caption{Experimental (Exp.) nickel L\textsubscript{3,2}-edge XAS spectra and calculated spectra from the charge-transfer multiplet model by Crispy software\cite{retegan_crispy}. The top of the figure shows the experimental XAS data for the exfoliated NiPS\textsubscript{3} sample with TEY signal (blue) and TFY signal (red). The subsequent spectra from top to bottom represent the calculated spectra corresponding to $\Delta$ decreasing from 5.0 eV to -1.0 eV.}
\label{fig:5}
\end{figure}

To determine the $\Delta$ of NiPS\textsubscript{3}, we performed theoretical calculations using the charge-transfer multiplet model. The comparison between the experimental data for the exfoliated sample and the theoretical calculation is presented in FIG.~\ref{fig:5}. From the calculated results, it can be observed that with the increase in $\Delta$, the most noticeable change is that the Ni L\textsubscript{3} and L\textsubscript{2} absorption edges transition from a single peak structure to a double peak structure, accompanied by a shift in the peak positions and changes in the satellite peaks. This trend is consistent with previous calculations\cite{kim2018charge,kang2020coherent,yan2021correlations}. When determining the $\Delta$ of NiPS\textsubscript{3} by comparing the features with calculations, however, the main focus and point of contention has been around the L\textsubscript{3} edge. By contrast, the most noticeable feature here, which differs from previous studies, is that the L\textsubscript{2} peak also splits into a double peak structure. The best match to the experimental results is the calculated spectra with $\Delta$=3 eV, yet the spectra with $\Delta$=2.5 eV and $\Delta$=3.5 eV also match some features of the experimental results very well. Therefore, we have obtained a relatively large $\Delta$ for NiPS\textsubscript{3}, suggesting it is highly likely that NiPS\textsubscript{3} falls within the regime of a Mott-Hubbard insulator.

\begin{figure}[htb]
\centering
\includegraphics[width=8.5cm]{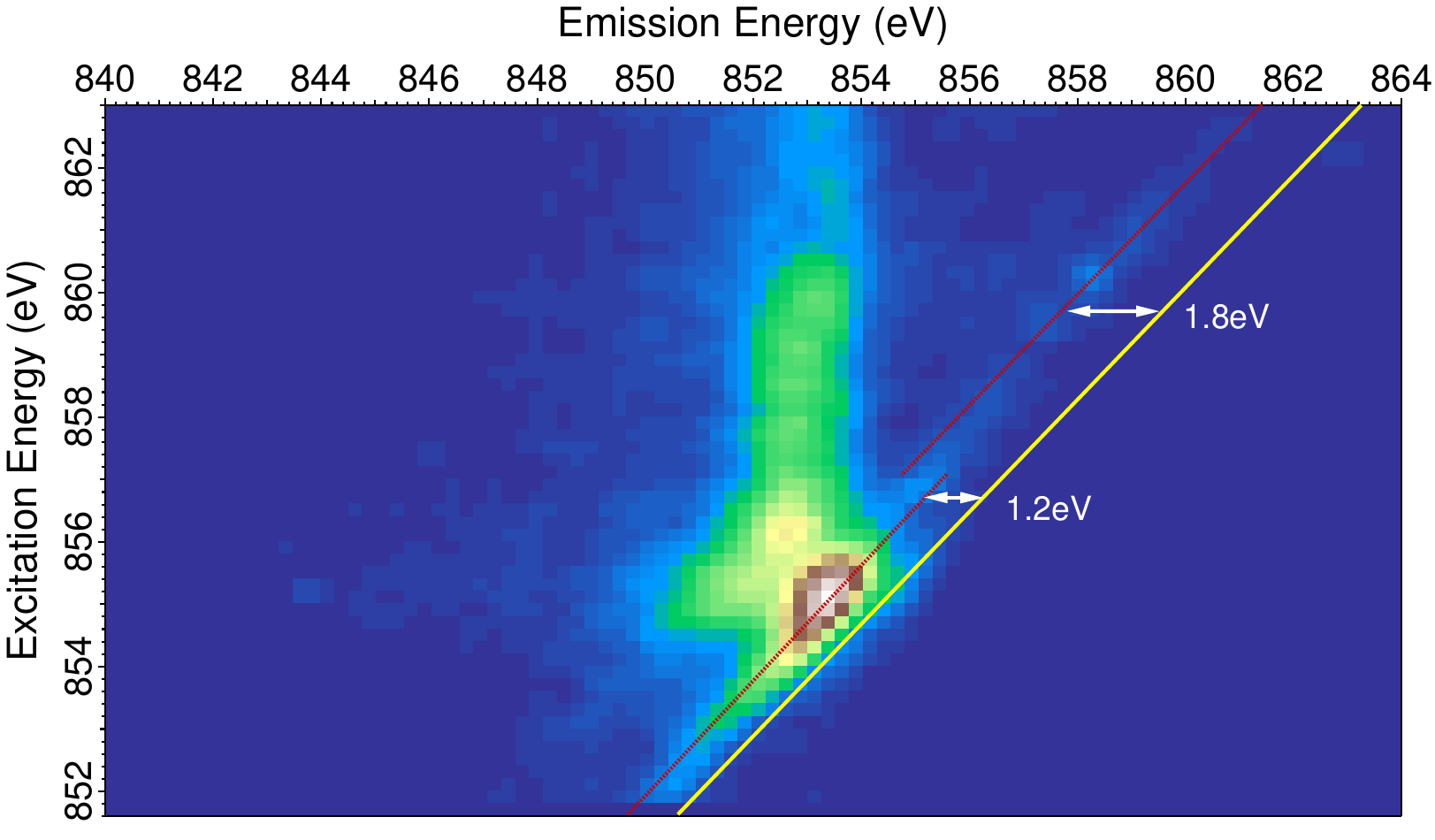}
\caption{Nickel L\textsubscript{3}-edge mapping of RIXS of NiPS\textsubscript{3}. Yellow solid line: elastic peak. Red dashed line: energy lost features.}
\label{fig:6}
\end{figure}

Thus far, we have determined the $\Delta$ of NiPS\textsubscript{3} with XAS and theoretical calculation. To further confirm that such a $\Delta$ categorizes NiPS\textsubscript{3} as a Mott-Hubbard insulator, we employed mapping of RIXS as shown in FIG. \ref{fig:6}. It can be observed that parallel to the elastic peak (yellow solid line), two resonant X-ray Raman features (red dashed line) were found with energy-loss of 1.2 eV and 1.8 eV. Based on the scale of energy-loss\cite{ament2011resonant} and previous optical spectra results\cite{kim2018charge}, both resonant X-ray Raman features were assigned to $dd$ excitations. Therefore, we found that the $dd$ excitations are dominant, with the charge-transfer features being too weak to be observed. This indicates that standard Hubbard model applies to NiPS\textsubscript{3}, with transitions occurring only among $d$-electrons themselves, rather than the $p-d$ model as in the case of charge-transfer insulator. Moreover, these energy-loss features match the previous RIXS result\cite{He2024}, except that we did not find the ultrasharp PL peak due to the limitation of resolving power. With the aid of calculations, they determined the expectation values describing the NiPS\textsubscript{3} wavefunction and revealed that NiPS\textsubscript{3} has a dominant Hund's character with a ground state $d^8$, instead of $d^9\underline L$ as in the Zhang-Rice scenario. This further proves that NiPS\textsubscript{3} belongs to a Mott-Hubbard insulator rather than a charge-transfer insulator.

\section{Conclusion}
In this study, we used XAS and RIXS to investigate the electronic structure of NiPS\textsubscript{3}. By comparing experimental data from as-grown and exfoliated samples with theoretical calculations, we found that NiPS\textsubscript{3} has a relatively large $\Delta$ (about 3 eV), indicating it is a Mott-Hubbard insulator. Additionally, we observed surface oxidation on as-grown samples, which may have contributed to the previous debates over the XAS results and was mitigated by exfoliation. RIXS data showed dominant $dd$ excitations with weak charge-transfer features, further supporting this classification. These findings resolve previous controversies, contribute to the broader knowledge of NiPS\textsubscript{3}, and emphasize the importance of controlling surface adsorption in studies of the electronic structure of vdW materials. 

\section{Acknowledgement}
This research used resources of the Advanced Light Source, which is a Department of Energy (DOE) Office of Science User Facility under contract no. DE-AC02-05CH11231. Work done by Q.T. and X.L. was supported by National Science Foundation (NSF) under Grant No. 2216008 and No. 1945364, and the U.S. DOE, Office of Science, Basic Energy Science (BES) under Award Number DE-SC0021064. Q.T. also acknowledges the support of the Laursen Fellowship at Boston University.

\bibliography{main}

\section{Figures}

\end{document}